\documentclass[12pt]{article}
\usepackage{amsmath,graphicx}
\usepackage{hyperref}
\usepackage{caption}
\title{Last but not Least: Additional Positional Effects on Citation and Readership in arXiv}
\author{
Asif-ul Haque\footnote{asif@cs.cornell.edu}, Paul Ginsparg\footnote{ginsparg@cornell.edu}\\
Dept of Information Science, Cornell University, Ithaca, NY 14853\\
}
\date{}

\begin{document}
\baselineskip 16pt plus 1pt minus 2pt

\maketitle

\begin{abstract}
We continue investigation of the effect of position in announcements of  newly received articles, a single day artifact, with citations received over the course of ensuing years.  Earlier work \cite{Dietrich,HG09} focused on the ``visibility'' effect for positions near the beginnings of announcements, and on the ``self-promotion'' effect associated to authors intentionally aiming for these positions, with both found correlated to a later enhanced citation rate.  Here we consider a ``reverse-visibility'' effect for positions near the ends of announcements, and on a ``procrastination'' effect associated to submissions made within the 20 minute period just before the daily deadline.
For two large subcommunities of theoretical high energy physics, we find a clear ``reverse-visibility'' effect, in which articles near the ends of the lists receive a boost in both short-term readership and long-term citations, almost comparable in size to the ``visibility'' effect documented earlier.  For one of those subcommunities, we find an additional ``procrastination'' effect, in which last position articles submitted shortly before the deadline have an even higher citation rate than those that land more accidentally in that position.   We consider and eliminate geographic effects as responsible for the above, and speculate on other possible causes, including ``oblivious'' and ``nightowl'' effects.

\end{abstract}

\section{Introduction}

In \cite{HG09}, we considered a surprising correlation between article position in the initial announcements of new articles and later citation impact.
As first described in \cite{Dietrich}, articles that appeared at or near the beginnings of the simultaneous web and e-mail announcements, appearing every weekday, received substantially higher median citations due to a combination of ``self-promotion'' and ``visibility'' effects.  ``Self-promotion'' reflected the tendency of some submitters to aim their submissions for just after the deadline, when they were most likely to appear near the beginning of the next day's announcements. ``Visibility'' reflected the extent to which articles that serendipitously appeared near the beginning,  by no conscious intent on the part of the submitter, nonetheless collected more median citations than had they appeared lower down in the listings.  As emphasized in \cite{HG09}, it is not immediately intuitive that position in the daily announcement of newly received submissions, a one-day artifact leaving no trace afterwards, could nonetheless leave its mark in long-term citation counts, accumulated years later. 

While  \cite{HG09} was in preparation, a local physics grad student suggested that some submitters might instead sometimes aim for the 
{\em bottom\/} of the list. One reason intimated for doing so was that some readers use a different URL from the website to track the newly announced submissions. In addition to the http://arXiv.org/list/$\ldots$/new URL\footnote{described in \cite{HG09}, where ``$\ldots$" denotes a specific subject class, e.g., astro-ph, hep-ph, hep-th}, there is an additional URL of the form http://arxiv.org/list/$\ldots$/recent, also prominently linked from the homepage, which collects the previous five days of announced articles. These are separated by day, with the most recent day at the top, so it was natural to present articles {\em within\/} each day as well in reverse order, in order that the numbering be continuous through the day boundaries. The other feature of the ``recent'' URL is that it displays Title and Author information, with only a link to the full abstract, so that many more entries appear within a browser window before paging down.  Following the appropriate links to read only those abstracts with relevant sounding titles may avoid some of the ``fatigue" that contributes to the ``visibility effect."
 In any event, the logic was that if enough readers access the announcements via the latter URL, and submitters are aware of this, then there would be an incentive  to aim submissions for the bottom of the list, as an alternative distinguished position.  

But what fraction of readers view the daily listings on the pages with reversed order? During the 2002--2004 period studied in \cite{HG09}, the webserver logs indicate that just over 80\% of the hep-th (``High Energy Physics --- Theory") and hep-ph  (``High Energy Physics --- Phenomenology") readers, and roughly 75\% of the astro-ph 
(Astrophysics) readers access the announcements via the /list/$\ldots/$new URL, and hence read in the canonical order.  An even greater number of readers continued to receive the announcements via the legacy e-mail subscription\footnote{roughly 1500 for each of hep-th and hep-ph, and about 3500 for astro-ph --- this was the timeframe during which the primary usage began to migrate to the web interface. Use of RSS feeds had not yet become widespread.}, and hence the vast majority of practitioners of these three subject areas read the announcements in the standard (forward) order.  
``Reverse-visibility'' effects were consequently not pursued in \cite{HG09}.
(Interestingly, the pattern was very different in some other subject areas, with roughly 75\% of mathematics, 85\% of computer science, and 50\% of condensed matter web interface users reading the arXiv daily listings in reverse order, via the  /list/$\ldots/$recent URLs.)

\begin{figure}[h!]
\includegraphics[scale=0.65]{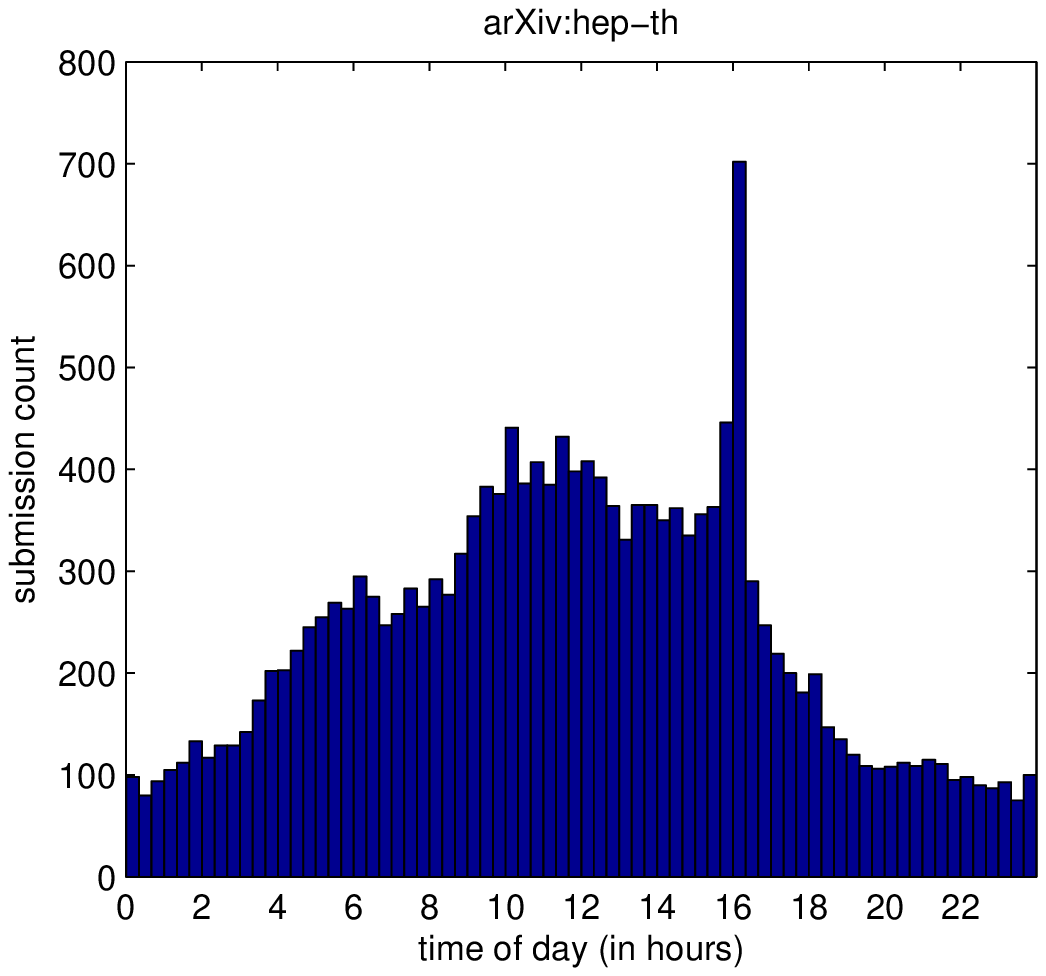}
\quad
\includegraphics[scale=0.65]{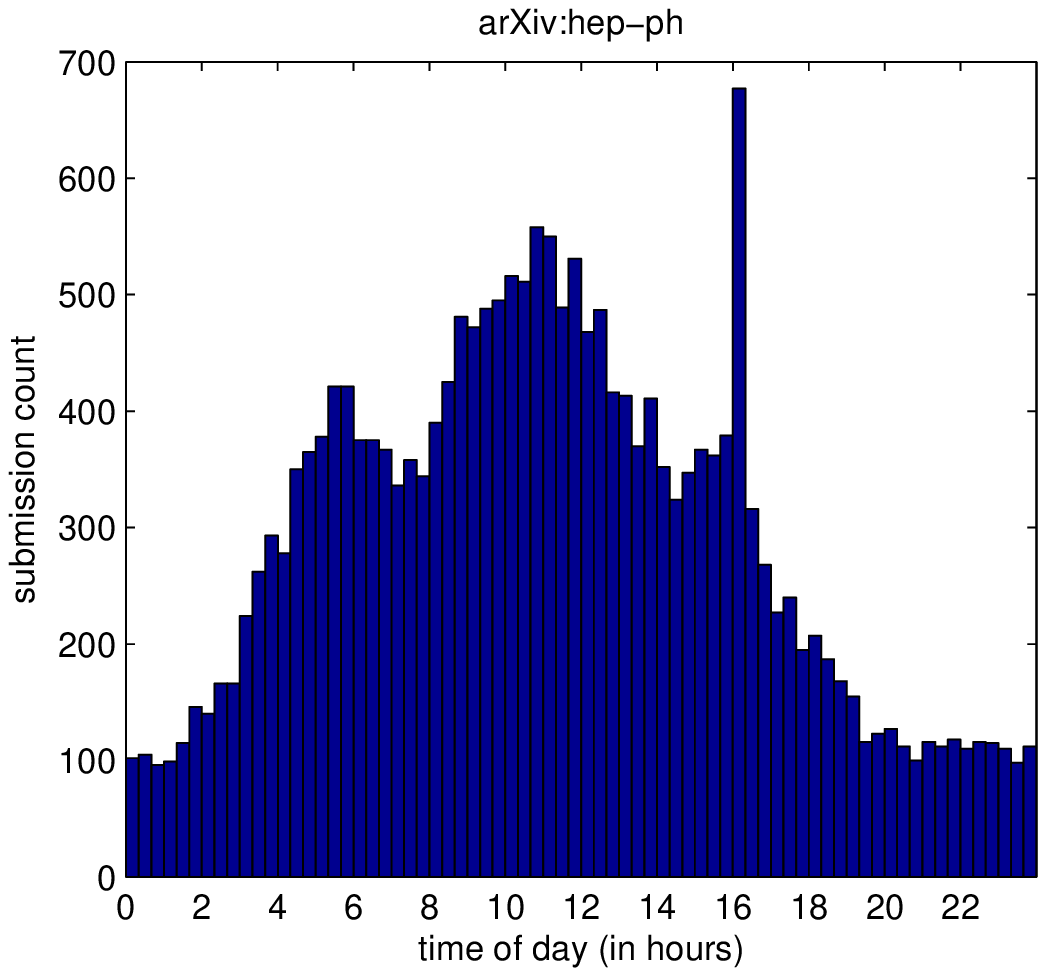}
\centerline{(a)\hskip3in(b)}
\caption{\small Figure 6 from \cite{HG09}. Number of (a) hep-th and (b) hep-ph submissions by time of day,\
 in 20 minute bins, during the period Jan 2002 -- Mar 2007.}
\label{fig:hepsubhist}
\end{figure}

Here we return to the ``reverse-visibility'' issue more systematically, and find that there is nonetheless a statistically significant enhancement of both readership and long-term citation behavior for submissions appearing near the end of the daily announcements.
That both the top and bottom of lists can be distinguished positions is familiar from cognitive studies, in which people asked to recall a list of items tend to recall most easily items near the beginning (primacy effect) and toward the end (recency effect) of the list \cite{Ebb13}.  
This ``serial position effect'' remains a very active area of research in cognition and memory, and is likely due to a form of interference, in which items cognitively processed in the middle of a list receive confounding interference from those on both sides, whereas those at the beginning or end receive less interference, only from succeeding or preceding items.

Choosing whether or not to follow a link from an abstract to retrieve a full text is a different exercise than recall of a randomly presented list, but nonetheless similar mechanisms may be at play.
For example, some readers may occasionally find sifting through the daily announcements to be something of a chore, paying more attention in the beginning, scanning more quickly through the middle, only to concentrate again towards the end for some sense of closure.
The task here, however, permitting real-time actions to be taken by readers while scanning the list (e.g., retrieving further information, as measured through full-text readership data), and as well permitting reconsultation of the list throughout the day, is sufficiently different that we maintain the visibility-related terminology used in \cite{Dietrich,HG09}.
An additional twist is that submitter behavior can affect placement of items within the next day's list, leading to the notions of ``willful primacy'' or self-promotion, 
and of ``willful recency'' or procrastination.

Another suggested motivation for posting near the 16:00 U.S.\ eastern time deadline is simply that it is a deadline: 
submitters with neither interest nor motivation to aim for the beginning of the following day's listings may nonetheless have a strong desire for their article to appear that night, perhaps to stake a precedence claim in a fast-moving field, or perhaps to offload a psychological weight.  For various reasons, including other commitments throughout the day carried by the busier and potentially higher profile researchers, or interactions with co-authors, etc., the submitter may delay posting until very close to the deadline. The deadline itself can even serve as the motivation for a final burst of focused activity to finish the text, rather than let it linger to the next day ad infinitum.
Indeed in fig.~\ref{fig:hepsubhist}, we have reproduced for convenience Fig.~6 from \cite{HG09}, showing the number of hep-th and hep-ph submissions received in 20 minute bins throughout the day.  While the largest spike appears in the 20 minute bin just after the 16:00 deadline, there is also evidence for a smaller burst in submissions leading up to that time.\footnote{Note that those are not typically submissions piling up in the final seconds before 16:00, which might instead indicate a premature submission effect: submitters with bad aim due to poor clock synchronization. While most such submitters are probably not aiming for the final position, such delayed submission behavior has the same effect.}  We refer to this effect  as the ``procrastination effect'', a flip-side of the ``self-promotion'' effect which, as we shall see,  shares some of its essential features.  


\section{Citation and Readership data vs.\ position}

\begin{figure}[h!]
\hskip-60pt\includegraphics[scale=0.5]{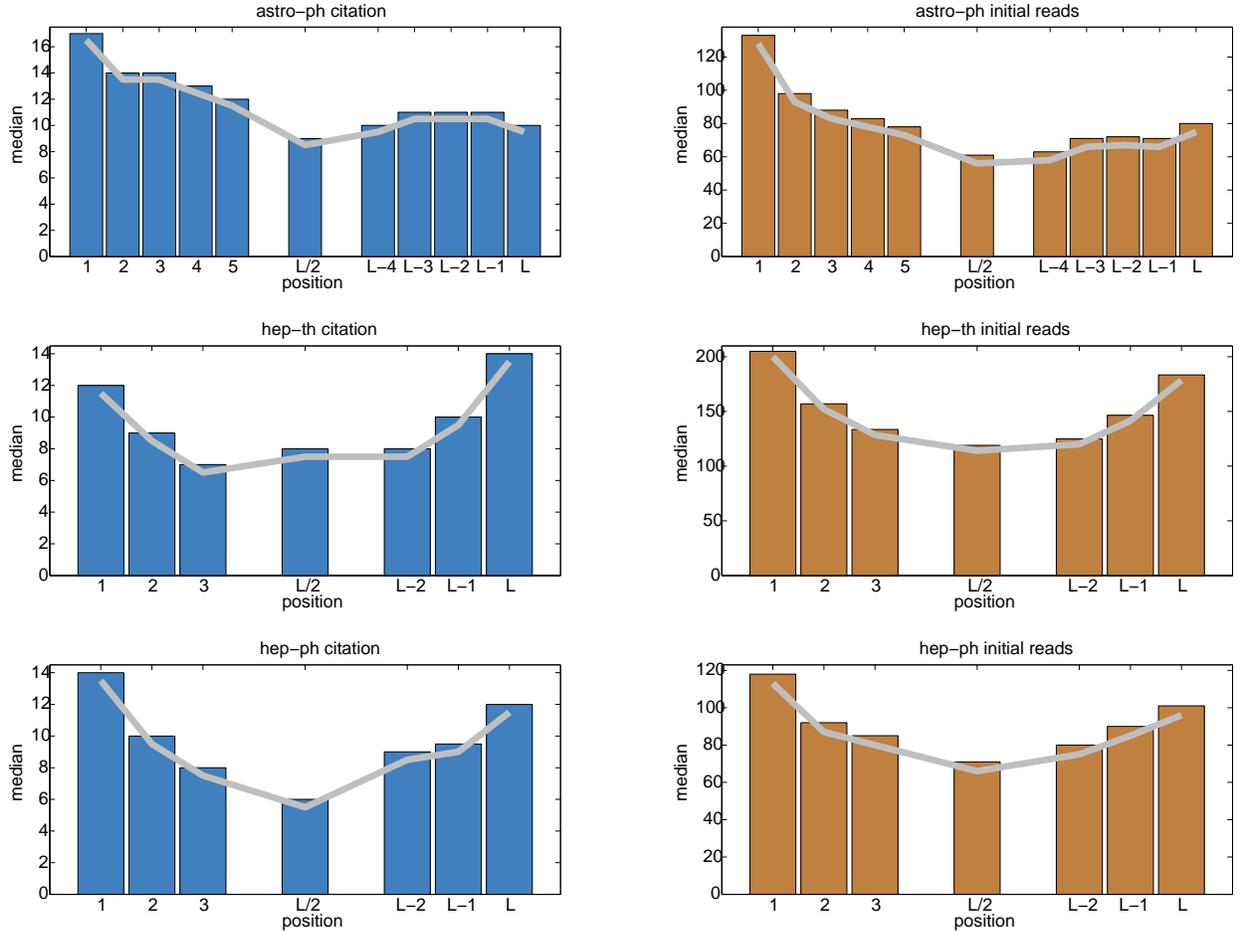}
\caption{\small 
Citation and readership statistics for the period 2002--2004 (same dataset as used in \cite{HG09}). L denotes the mailing length: L is the last article, L-1 is the second-to-last, L/2 is the middle, and so on.
Mailings shorter than 11 were ignored for astro-ph (requiring at least five articles before and after the middle), and similarly mailings shorter than 7 were ignored for
hep-th and hep-ph (requiring at least three before and after middle). Of the initial total of 776 mailings, this left roughly 770 mailings for astro-ph, 720 for hep-th and 750 for hep-ph, and the average mailing size L remained 30, 13, and 16, respectively, for that period.
The number of articles in each position is the same as the number of mailings, except  the middle position has roughly 1.5 times as many articles as mailings since both middle positions are kept for L even.  The readership data is for the initial active  
period of 10, 25 and 15 days respectively for astro-ph, hep-th and hep-ph, as explained in \cite{HG09}.}
\label{fig:posstats}
\end{figure}

In order to assess any systematic effects for articles appearing near the ends of the mailings, we present in fig.~\ref{fig:posstats} three sets of articles for any mailing of size L.
The articles near the beginning of the announcement are labelled according to position 1,2,3,$\ldots$, those near the end of the announcement are labelled $\ldots$,L-2,L-1,L, and the one (or two, for L even) in the middle is labelled L/2.
Isolating the submissions in reverse order near the end of the announcements  permits identifying any coherent end effects, independent of the varying length L.  The underlying dataset is the same as used in  \cite{HG09}, except that only mailings above a minimum size threshold have been retained, as described in the figure caption.

We see visual evidence in fig.~\ref{fig:posstats} for a ``reverse-visibility'' effect near the end of the hep-th and hep-ph mailings, both in readership during the initial active periods of the first few weeks after announcement, and as well in median citations received years afterwards. The effect is most prominent in the median citations received by articles near the end of the hep-th mailings (middle figure on left), coinciding with the noted pile-up in submissions just {\it before\/} the 16:00 deadline,
reproduced here in fig.~\ref{fig:hepsubhist}a.  Overall the full-text readership data confirms that the majority of readers peruse the lists in the standard forward order, with the greatest number of accesses to articles in the first few announcement positions.

\begin{figure}[h!]
\hskip-60pt\includegraphics[scale=0.75]{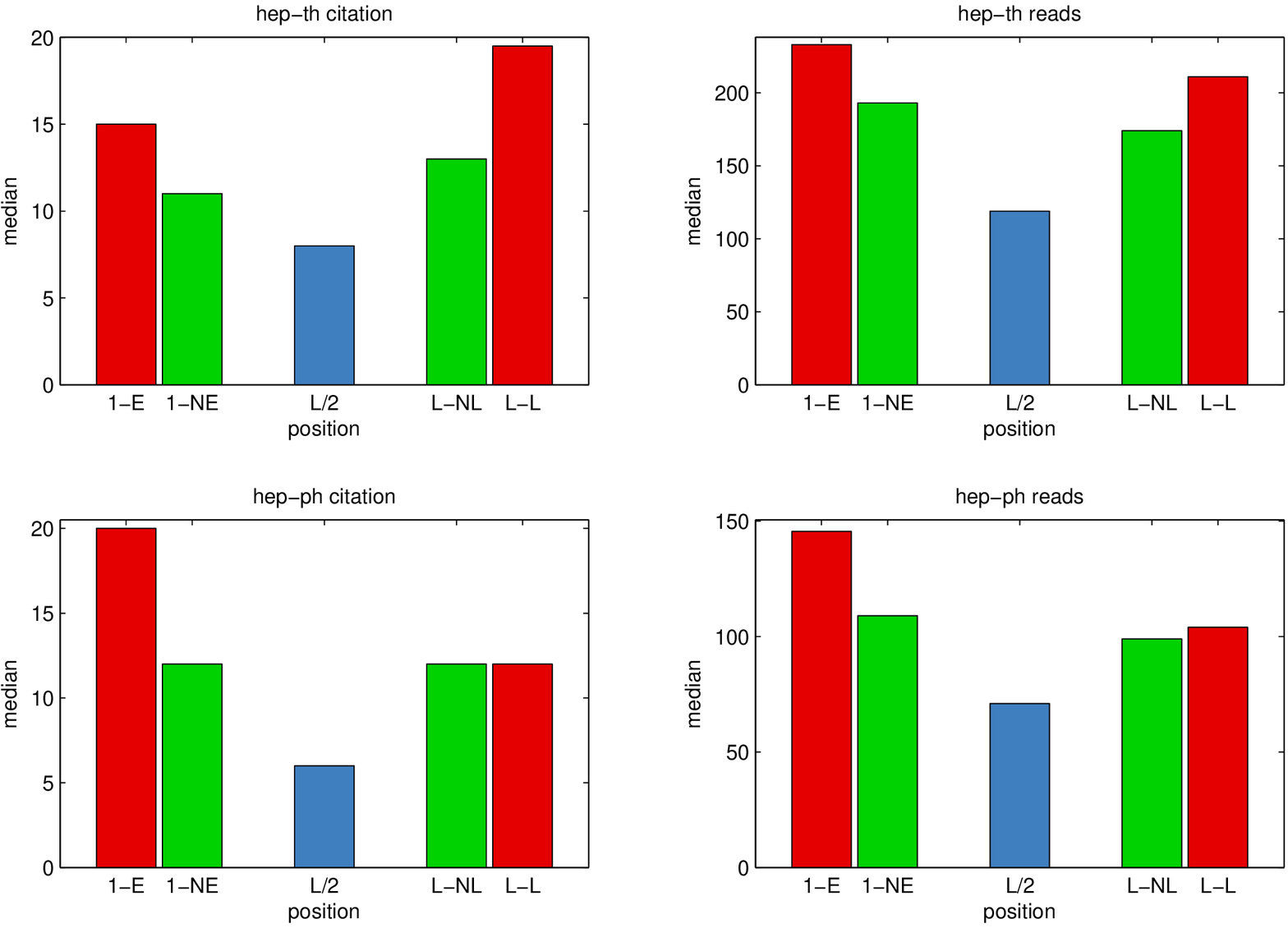}
\caption{\small 
Citation and read plots for hep-th and hep-ph isolating early and late contributions. Here 1-E and 1-NE denote  first position ``early'' and ``not early'', L/2 denotes the middle position, L-NL and L-L denote last position ``not late'' and ``late'' submissions, respectively.}
\label{fig:hepfls}
\end{figure}

The effect is less dramatic for the (longer) astro-ph announcements, so we restrict attention in what follows to the hep-th and hep-ph subject areas. As an aside, these two disciplines are very closely related, having emerged from the same theoretical particle theory research community following the re-emergence of string theory in the mid 1980's, with the more mathematically inclined represented in hep-th, and the more phenomenologically oriented in hep-ph, and a small subset of researchers contributing to both.

For further clarification of the effects in fig.~\ref{fig:posstats},  we separate out in  fig.~\ref{fig:hepfls} the contributions to the first position according to whether the articles were received ``early'' (E), within the first 20 minutes after the 16:00 eastern time deadline, and the contributions to the last position according to whether the articles were received ``late'' (L), within the last 20 minutes before the next 16:00 eastern time deadline.
By this criterion, 32\% and 37\% of the first position submissions to hep-th and hep-ph, respectively, were early, 
and 22\% of the last position submissions to each were late.

In principle, one might expect no difference between either 1-NE or L-NL submissions and L/2 submissions, since all were submitted far enough from the deadline to be insulated from ``self-promotion'' or ``procrastination'' effects.  Yet the median citation differences between submissions appearing at L/2 and the other two positions are significant at the 1\% level\footnote{We have used the non-parametric Mann-Whitney U (MWU) and Kolmogorov-Smirnov (KS) tests.}. The difference between the 1-NE and L/2 citations is the visibility effect discussed in \cite{Dietrich,HG09}, and the difference between the L-NL and L/2 citations is part of the new ``reverse-visibility'' effect.
This latter is likely some combination resulting from the smaller percentage of readers who access the lists in reverse order via the /list/$\ldots$/recent URLs, from readers whose attention lapses in the middle of lists to refocus when the end is in sight, and from readers who may consciously or subconsciously be expecting to finding higher quality articles near the end, due to the next effect we discuss.

The distinction between 1-E and 1-NE submissions is attributed to ``self-promotion'', i.e., the 1-E articles are distinguished by having been intentionally targeted by submitters to appear early in the announcements.  The distinction between the  L-NL and L-L articles is what we've termed ``procrastination'', with the latter articles having been more likely to appear in the last position due to submission within the final 20 minutes before deadline. (We note that ``procrastination'' might involve a slightly different mentality in Europe, where the deadline occurs in the late evening rather than during the working day.)
For hep-th, the median citation differences between submissions in the 1-E and 1-NE positions,  and between
those in the L-NL and L-L positions in fig.~\ref{fig:hepfls} are significant at the 1\% level.
While the L-L submissions appear to have a strikingly higher median citation rate of 19.5 compared to the self-promotion enhanced 1-E rate of 15, the 4.5 citation
 difference is not significant even at the 10\% level by MWU test (P=0.1138 or only at the 11.38\% level), so the 1-E and L-L positions should be considered as statistically similar.
 
For hep-ph, all of the 1-NE, L-NL, and L-L positions in fig.~\ref{fig:hepfls} have 12 median citations, while 1-E submissions have a median of 20 (the enhancement corresponding to the previously noted self-promotion effect).
Only hep-th appears to have the ``procrastination'' effect, in which submissions made just before the deadline tend to receive more citations.  It is not entirely clear why two such similar disciplines would exhibit this distinction --- perhaps practitioners of the less experimentally oriented discipline, operating on a shorter timescale and hence feeling more competition, perceive more of a need to stake their last minute precedence claims to avoid being scooped.

\section{Geographic/Timezone effects}

The other curious feature of  figs.~\ref{fig:posstats},\ref{fig:hepfls} is the dip in the middle, i.e., the lower median citation rate
for articles appearing in the vicinity of position L/2.
This raises the question of whether submissions that appear in the middle of the announcement lists are subject to some other systematic bias, such as geographic.  Suppose that researchers located in timezones whose workday is far displaced from 16:00 U.S.\ eastern time
also happen to receive systematically lower citations, for some related or unrelated reason. Then a geographic bias would explain not only that 
salient feature of  figs.~\ref{fig:posstats},\ref{fig:hepfls}, it might also be partly responsible for what was identified as a visibility effect in \cite{Dietrich,HG09}.

The end-of-workday 16:00 U.S. eastern time deadline for submissions corresponds (during daylight savings time) to afternoon (13:00--16:00) in the continental U.S.\ (and to 10:00 and noon in Hawaii and Alaska, resp.), to late afternoon in South America (15:00--17:00), late evening in Western Europe, Middle Africa, and Scandinavia  (21:00-23:00), around midnight in the Middle East and Western Asia  (23:00--01:00),  
to middle of the night in Eastern Asia (01:00--05:00 for India, China, Korea, Japan), and to early morning in Australia and New Zealand (04:00--08:00).  The deadline corresponds to the middle of the night for much of Asia, and the natural end of the workday in those regions corresponds instead to the middle of the daily submission period, causing submissons to land closer to the middle of the daily announcements.

\begin{figure}[h!]
\begin{center}
\includegraphics[scale=0.6]{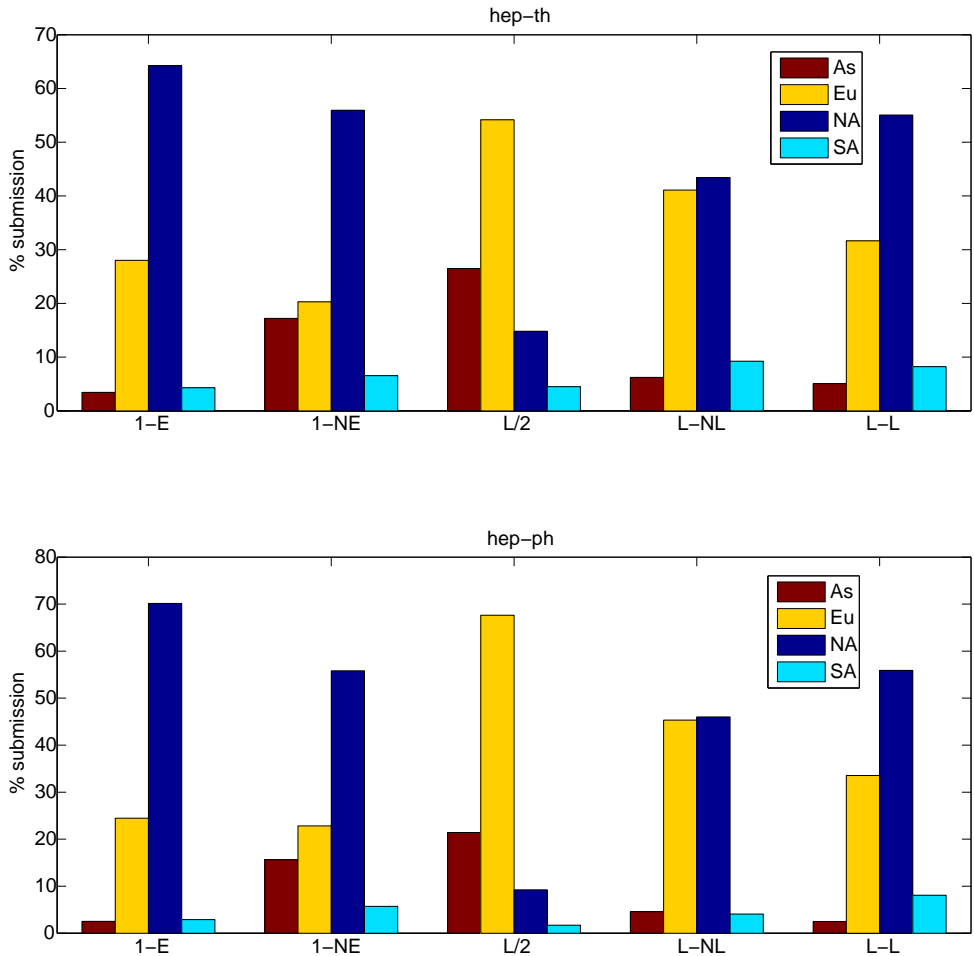}
\end{center}
\caption{\small 
Geographical distribution of 1-E, 1-NE, L/2, L-NL, L-L articles for hep-th and hep-ph, using the country specified by the domain of the submitter's e-mail address.
North America (NA) is out of phase with Asia (As), although a few dedicated submitters from Asia submit early/late. 
A substantial fraction of the middle articles are from Europe (Eu). Contributions from South America (SA) are under 10\% for each bin.}
\label{fig:heppos}
\end{figure}

To begin assessing whether the larger admixture of such submissions is responsible for the dip, fig.~\ref{fig:heppos} further subdivides the bins of fig.~\ref{fig:hepfls} according to the registered e-mail address of the submitter.\footnote{This is in rough correspondence with the geographic location of the submitter, with a few qualifications:
(i) .com and .org addresses were taken as US and hence North America. This should not be problematic since during the 2002--2004 submission period in question, only a very small percentage of submitters were using .com addresses and there should be no effect on the medians or statistical significances of differences.
(More care would be required for analysis of astro-ph, since the world's largest astronomy research facility, eso.org, is a European center.)
(ii) The e-mail address only corresponds to the location of the submitting author, so this form of classification ignores the effects of cross-continental collaborations. Our experience is that the choice of submitting author nonetheless typically reflects the majority point of origin.
(iii) The submitting author may no longer be physically at the same location as the e-mail address first used to create the registration. For the most part, however, submitters have kept their contact information up-to-date.
(iv) The boundary between Europe and Asia can be defined either politically or geographically. Here we use the political definition, including Russia with Europe.
(v) Asia is far from homogeneous, even familiar countries such as China, Japan, South Korea, Israel and India are at differing levels of
economic and hence academic development. The coarse aggregate measures used here accurately reflect these weighted heterogeneous distributions.}
The geographic distributions for the 1-E vs.\ 1-NE and L-NL vs.\ L-L submissions are clearly different.
The  middle L/2 position decomposes most distinctly from the distributions for first and last positions, with the majority at L/2 coming from Europe, and with a much larger percentage coming from Asia.

If we consider the analog of fig.~\ref{fig:hepsubhist}, but only for As submissions, then neither hep-th nor hep-ph show spikes at the 16:00 deadline. The overall submission pattern is as expected from fig.~\ref{fig:heppos}, out of phase with NA: with a smooth peak in the 4:00 range (early in the morning on the East coast), about 7 hours before the 11:00 aggregate peaks in fig.~\ref{fig:hepsubhist}.  We have verified the geographic locations of the internet IP addresses used to upload each of the 1-E submissions from submitters registered with As e-mail addresses, and the majority did indeed arrive from middle-of-the-night timezones.
The small remainder came from As registered users temporarily operating from NA or Eu locations, as temporary visitors or attending summer workshops or schools.
(Curiously, the submission pattern for As submissions to astro-ph does show a significant spike at the 16:00 deadline, so self-promotion in astro-ph is sufficiently attractive to entice submitters to operate at inconvenient hours.  The roughly 100 such submissions over the greater than 5 year time period nonetheless corresponds to a low rate, of only one per every 2.5 weeks.  About 75\% of those did arrive from internet IP addresses in As timezones, mainly from Israel, Japan, and India --- at local times ranging from 23:00 to 5:00.) 






\begin{table}[h]
\begin{center}
\begin{tabular}[c]{| c | c | c | c | c | c |}
\hline
& Eu & NA & As & SA &EuNA\\
\hline
astro-ph &        9 &    12 &  7&  6 &10\\
hep-th     &        8  &   13 &  7& 4 &9\\
hep-ph     &       7 &    12  &  6& 5 &10\\
\hline
\end{tabular}
\hskip.5truein
\begin{tabular}[c]{| c | c | c | c | c | c |}
\hline
& Eu & NA &As & SA \\
\hline
astro-ph & 10523 &  9272 &      2296  &   452\\
hep-th     &  4495  &  2688 &      2030  &   606\\
hep-ph    &   6537  &  3100 &       2220&   322\\
\hline
\end{tabular}
\caption{\small (a) table of median citations of 02-04 articles. All pairwise differences are significant at the 5\% level.  (b) number of submissions from each continent.}
\label{table:medcit}
\end{center}
\vskip-15pt
\end{table}
Table~\ref{table:medcit} shows the geographic citation differences for submissions associated to the 4 most active continents, independent of announcement position
(i.e., aggregated over all submissions from each of  the four regions during the 2002--2004 timeframe).
The median citations of 7, 7 and 6 of Asia (As) articles for astro-ph, hep-th and hep-ph, respectively, are about  30\% less than the median citations  of 
10, 9 and 10 for the same subject areas from the combination of Europe and North America (EuNA).

We now assess to what extent geographic/timezone effects,  causing those Asian submissions with fewer citations to be disproportionately represented among the middle submissions,  might account for the dip in the middle of figs.~\ref{fig:posstats},\ref{fig:hepfls}.
In the following, statistical significance limits are set at the 5\% level, and we  ignore South America.

\begin{table}[h]
\begin{center}
\begin{tabular}[c]{| c | c | c | c | c | c |}
\hline
hep-th& 1-NE & L/2\\
\hline
EuNAAsSA & 11 & 8 \\
EuNA & 12 & 8 \\
EuNAAs & 11 & 8 \\
NA & 12 & 11 \\ 
Eu & 12 & 8 \\ 
As & 8 & 8 \\ 
\hline
\end{tabular}
\hskip1truein
\begin{tabular}[c]{| c | c | c | c | c | c |}
\hline
hep-ph& 1-NE & L/2\\
\hline
EuNAAsSA & 12 & 6 \\
EuNA & 13 & 6 \\
EuNAAs & 12 & 6 \\
NA & 14 & 8 \\ 
Eu & 12 & 6 \\ 
As & 9 & 7 \\ 
\hline
\end{tabular}
\caption{\small Median citations to hep-th and hep-ph submissions at positions 1-NE and L/2, for various subsets of geographic regions.
The data from these bins succinctly captures any geographic bias and visibility effects.}
\label{table:hepgeo}
\end{center}
\vskip-15pt
\end{table}

First we compare citations of European (Eu) and North American (NA) submissions.
For the five subsets 1-E, 1-NE, L/2, L-NL and L-L in fig.~\ref{fig:hepfls}, the differences between NA and Eu median citations are not statistically different, 
except for the hep-ph L/2 submissions. In that bin, NA submissions have a median of 8 citations while Eu submissions have a median of 6
(and even there only significant by the MWU test, not by the KS test --- by contrast, the corresponding medians of 11 and 8 for hep-th  at L/2 are not significant by either test at the 5\% level).
Overall there is little evidence of geographic bias between Eu and NA submissions, in accord with the observation in \cite{Dietrich}.\footnote{As described in \cite{HG09}, the cuts in \cite{Dietrich} resulted in focusing on higher cited articles with potentially attenuated geographic bias, so it is worthwhile to confirm for the larger sample.}

Next we include Asian (As) submissions and compare citations of the combined EuNAAs to those of the combined EuNA submissions.
The first rows  in Table~\ref{table:hepgeo} show the median citations for the combined submissions from the four continents
(EuNAAsSA) for articles in positions  1-NE and L/2 (just two of the bins from fig.~\ref{fig:hepfls}), for hep-th and hep-ph.
Restricting to just the combined EuNA submissions in the second rows, only the hep-ph 1-NE median has a slight statistically significant increase (from 12 to 13 in the first column of the table on the right).
Reincluding the As submissions in the third row, we see that the median citation for articles in the  L/2 position is unaffected, and
consequently those are {\em not\/} responsible for the smaller median for articles in that position.
The median of 8 for As submissions to hep-th at the L/2 position is actually the same as that for EuNA 
(none of As,  Eu and NA have a statistically significant difference at that position).
The As L/2 median of 7  for hep-ph  is even slightly larger than the EuNA L/2 median of 6
(the small median advantage for NA submissions over those from Eu and As at L/2 is not statistically significant, 
with EuAs constituting over 80\% of the submissions at that position).  
Since the Asian submissions are concentrated near the middle of the mailings and receive similar median citations as  Eu or NA submissions at
L/2, they do not add a geographic bias to the positional effects analyzed in \cite{HG09}.

By contrast, there is a drop by one median citation at the 1-NE position between rows 2 and 3 of table~\ref{table:hepgeo},
confirming that adding the lower median citations of As submissions to those from EuNA do lower the median for positions other than L/2 (though
with only about 80 As submissions at that position, the signal is weak, i.e., significant by only one of the two tests, at the 5\% level).
A similar comparison between EuNA and EuNAAs submissions at the 1-E and L-L positions is not possible since there are even fewer As submissions (under 10) at those positions.





We close here with some additional comments about data in the tables. 
It might seem odd at first sight that the overall median NA hep-th citation in table~\ref{table:medcit} is 13, while the corresponding
NA median hep-th citations at the 1-NE and L/2 positions in table~\ref{table:hepgeo}  are 12 and 11, but the former is pulled up by other positions, 
including a 1-E median citation of 16.
In table~\ref{table:hepgeo}, the majority of the 1-NE vs.\ L/2 effect in hep-th median citations arises from Eu submissions.
Since 1-NE submissions from Eu are typically submitted long after ordinary working hours, 
this difference suggests an additional ``nightowl'' effect (not yet further investigated), complementary to the ``procrastination'' effect, that researchers who habitually work late into the night receive more median citations.  It would be instructive to find further quantitative evidence that researchers with obsessive work habits (working to meet deadlines, or working through the night) ultimately have more impact.



In summary, the results of this section clarify the question posed by table~\ref{table:medcit}, suggesting that  any period with fewer NA submissions would have a geographically induced diminution of median citations. Even though NA articles in aggregate tend to receive more median citations, the specific subset that appears in the middle of mailings perform about the same as those from elsewhere.
This suggests yet another curiosity in aggregate citation behavior, an ``oblivious'' effect:
researchers who operate oblivious to deadlines, submitting neither shortly before nor shortly afterwards, tend to get fewer median citations!

\section{Discussion}

In \cite{HG09}, the effects of visibility and self-promotion were disentangled by comparing the subset of articles that serendipitously appeared in early positions (due to administrative moves or slow submission days) with those targeted to appear there.   An analogous procedure here would be to consider articles originally slated for final positions but  shifted to middle positions by administrative moves. A statistically significant lower median citation rate for this subset compared to those that remained in final positions would provide alternate confirmation of the ``reverse-visibility'' effect. Similarly, if articles submitted shortly before the deadline and administratively moved to middle positions continued to show a higher median citation rate than either middle or late position articles, this would provide alternate confirmation of the ``procrastination'' effect. 
While the medians in these cases did tend strongly in the expected directions, the sets in question were unfortunately not large enough to make statistically significant statements (even relaxing the definition of ``middle'' to be neither first three nor last three positions).

As discussed in \cite{HG09}, the effects uncovered here result from unintentional properties of the announcement system during the timeframe studied. There have since been some changes in the system, including subdivision of astro-ph into subcategories, but many of the positional effects remain possible in the current system.
Some users have suggested randomizing the daily order entirely, either uniformly for everyone, or individually for each user. Others have 
pointed out that might render an unintentional disservice to readers, who perhaps benefit from seeing self-promoted or procrastinated articles brought preferentially to their attention. More modern presentation systems have also been suggested, such as more subgrouping by topics or enhanced graphical representations (2d concept maps, etc.).
The most likely remediation of these issues remains some form of personalization system, in which preferences actively registered by users via controlled keywords or search terms, combined with passively collected past usage data (from the same user on an opt-in basis), provide user-specific highlighting or reordering of entries.
This would ultimately mitigate the global resonance phenomenon, unique to this resource, of entire research communities viewing the same material in the same order on a daily basis.

The citation effects analyzed here and in \cite{HG09} have been formulated in terms of the median, because the mean of these heavy-tailed distributions would be strongly affected by the highly cited articles in the tail. Those heavily cited ``elite'' articles are moreover less likely to be subject to the various visibility and related effects in the long-term.  These effects, however, do extend beyond the median, or typical, article, as seen in the upper quartile distributions of \cite{HG09}.
It is also important to note that the difference in read and citation rates as a function of list position is large, in some cases a factor of two,
and, for the most part, independent of the quality of the paper.  This has substantial implications for use of these metrics in assessing individuals and organizations.




Finally, since such intriguing differences in behavior between practitioners of such closely related disciplines (hep-th and hep-ph) are seen here, it will be informative to assess the behavioral characteristics of other disciplines within the arXiv dataset. 
Further insight can be obtained by tracking the behavior of individual (anonymized) readers in the usage logs.
It is also possible to consider the time dependence of these effects, using datasets after the 2002--2004 sample used here, now that their long-term citations have stabilized. 
Other short-term visibility-type effects on long-term citation rates can as well be investigated, including work underway on whether an article's lucky appearance in a smaller daily announcement list, or unlucky co-occurrence with an article destined to be highly cited, have measurable long-term citation effects.

\bigskip\bigskip
\noindent{\bf Acknowledgements:}
We thank James Zibin for an e-mail query that reminded us to revisit the reverse-visibility question, and thank the reviewers for suggesting helpful clarifications.

\newpage

\bibliographystyle{abbrv}
\bibliography{positionaleffects}

\end{document}